\documentclass[11pt]{revtex4}

\usepackage{subfigure}
\usepackage{epsfig}
\usepackage{amssymb,amsmath}

\begin{document}

\title{Agreement of Neutrino Deep Inelastic Scattering Data with Global Fits of Parton Distributions}

\author{Hannu Paukkunen}
\email{hannu.paukkunen@jyu.fi}
\affiliation{Department of Physics, University of Jyv\"askyl\"a, Finland \\ 
             Helsinki Institute of Physics, Finland}

\author{Carlos~A. Salgado}
\email{carlos.salgado@usc.es}
\affiliation{Departamento de F\'\i sica de Part\'\i culas and IGFAE, Universidade de Santiago de 
Compostela, Galicia--Spain}

\begin{abstract}
The compatibility of neutrino-nucleus deep inelastic scattering data within the universal, factorizable
nuclear parton distribution functions has been studied independently by several groups
in the past few years. The conclusions
are contradictory, ranging from a violation of the universality up to a good agreement, most of the
controversy originating from the use of the neutrino-nucleus data from the NuTeV Collaboration.
Here, we pay attention to non-negligible differences in the absolute normalization between different
neutrino data sets. We find that such variations are large enough to prevent a tensionless fit
to all data simultaneously and could therefore misleadingly point towards nonuniversal nuclear effects.
We propose a concrete method to deal with the absolute normalization and show that an agreement between
independent neutrino data sets is established.

\end{abstract}

\maketitle

A well-established procedure in nearly all phenomenological analyses of high-energy collisions involving
hadrons is the division of the cross sections in universal sets of parton distribution functions (PDFs)
$f_i(x,Q^2)$ and short distance partonic processes. Here, $x$ is the momentum variable, $i$ labels the
parton types, and $Q^2$ is the scale specific for the process. A theoretical foundation for such a
procedure is provided by the theorem of collinear factorization \cite{Collins:1989gx},
applicable to a wide range of hard (involving a large scale $Q^2 > 1 \,{\rm GeV}^2$) processes in
high-energy lepton+nucleon and nucleon+nucleon collisions. Although the non perturbative nature of 
the PDFs still prevents their precise calculation from the first principles of QCD, their scale dependence
is given by the well-known Dokshitzer-Gribov-Lipatov-Altrarelli-Parisi (DGLAP) equations \cite{Dokshitzer:sg,Gribov:ri,Gribov:rt,Altarelli:1977zs}
which resum the large logarithms $\sim \log(Q^2)$ emerging from collinear QCD radiation.
Ultimately, the validity of the factorization is verified in global analyses comparing a diverse set
of experimental cross sections to the PDF-dependent calculated values.
The initial conditions $f_i(x,Q_0^2)$ for the DGLAP evolution are
iteratively adjusted to see if a single set that can reproduce all the data exists. The PDFs and
their uncertainties extracted in this way provide an indispensable tool for estimating signals,
backgrounds and acceptances in other high-energy experiments. Clearly, the validity of
factorization is of utmost importance for the phenomenology of high-energy hadronic collisions.

The global analyses of the nuclear parton distribution functions (nPDFs) study the applicability
of the collinear factorization in hard processes involving bound nucleons. The most recent 
analyses  \cite{Eskola:2009uj,Hirai:2007sx,Schienbein:2009kk,deFlorian:2011fp} include
data on charged lepton nuclear deep inelastic scattering (DIS); Drell-Yan dilepton production
in proton-nucleus collisions; and, in some cases, production of high transverse momentum pions
in deuteron-gold collisions and neutrino-nucleus DIS. The good overall agreement with the available
high-energy data supports the existence of universal, process-independent nPDFs. The nPDFs find
applications in high-energy nucleus-nucleus collisions, playing an essential role e.g. in the
heavy-ion program of the LHC.

The adequacy of the factorization in nuclear environment is of importance also from the point of view 
of free nucleon analyses \cite{Martin:2009iq,Ball:2011uy,Nadolsky:2008zw}, which often wish to employ 
nuclear data as an additional constraint. One such
process is the neutrino-nucleus DIS, which is useful for constraining e.g. the strange quark distribution,
but the weakness of the neutrino interactions requires the use of a nuclear target. This process 
has recently invoked special attention as its compatibility within the framework of universal nPDFs was
questioned \cite{Schienbein:2009kk,Schienbein:2007fs}. It was even declared \cite{Kovarik:2010uv}
that there is no way to satisfactorily reproduce the neutrino-nucleus and the other nuclear data
simultaneously with a single set of nPDFs. This could have far-reaching consequences as the
inability to find a set of nPDFs which at the same time describes all the considered data 
is the expected sign e.g. of a violation of the universality, or a breakdown of the DGLAP evolution.
However, the same signal can also occur if one or more of the experimental data sets contains
unrecognized systematic inaccuracies.

Contradictory results were first presented in \cite{Paukkunen:2010hb}, where up-to-date nPDFs were
found to give an excellent overall agreement with neutrino data from CDHSW \cite{Berge:1989hr}, 
CHORUS \cite{Onengut:2005kv} and NuTeV \cite{Tzanov:2005kr} Collaborations, although issues with 
the normalization of the NuTeV data were identified possibly explaining the results
of \cite{Kovarik:2010uv}. Similar conclusions were reached in \cite{deFlorian:2011fp}
where data from all these experiments were utilized in a global nPDF analysis
without an apparent disagreement. However, the baseline
PDFs utilized there \cite{Martin:2009iq} were already constrained by the NuTeV DIS
data and their uncertainties were treated as additional, uncorrelated point-to-point errors.
Furthermore, the analysis did not use the absolute cross sections, but the far more scarce
structure function data. Given all this, the neutrino data did not carry as 
heavy an importance as in \cite{Kovarik:2010uv}. For more comprehensive review of
the present situation, see \cite{Kopeliovich:2012kw}.

In this Letter, we will show that when accounting for the overall normalization of the
experimental data in neutrino DIS, all three data sets do show a uniform pattern of nuclear modifications,
well reproduced by the existing nPDFs. This reinforces the conclusions of \cite{Paukkunen:2010hb}, in 
a model-independent way, supporting the functionality of the
factorization in neutrino DIS. We make the point even more concrete by employing a method based on
the Hessian error analysis to verify the consistency of these data with CTEQ6.6 \cite{Nadolsky:2008zw}
and EPS09 \cite{Eskola:2009uj} global fits.

\begin{figure*}[ht]
\begin{minipage}[b]{0.45\linewidth}
\centering
\includegraphics[width=\textwidth]{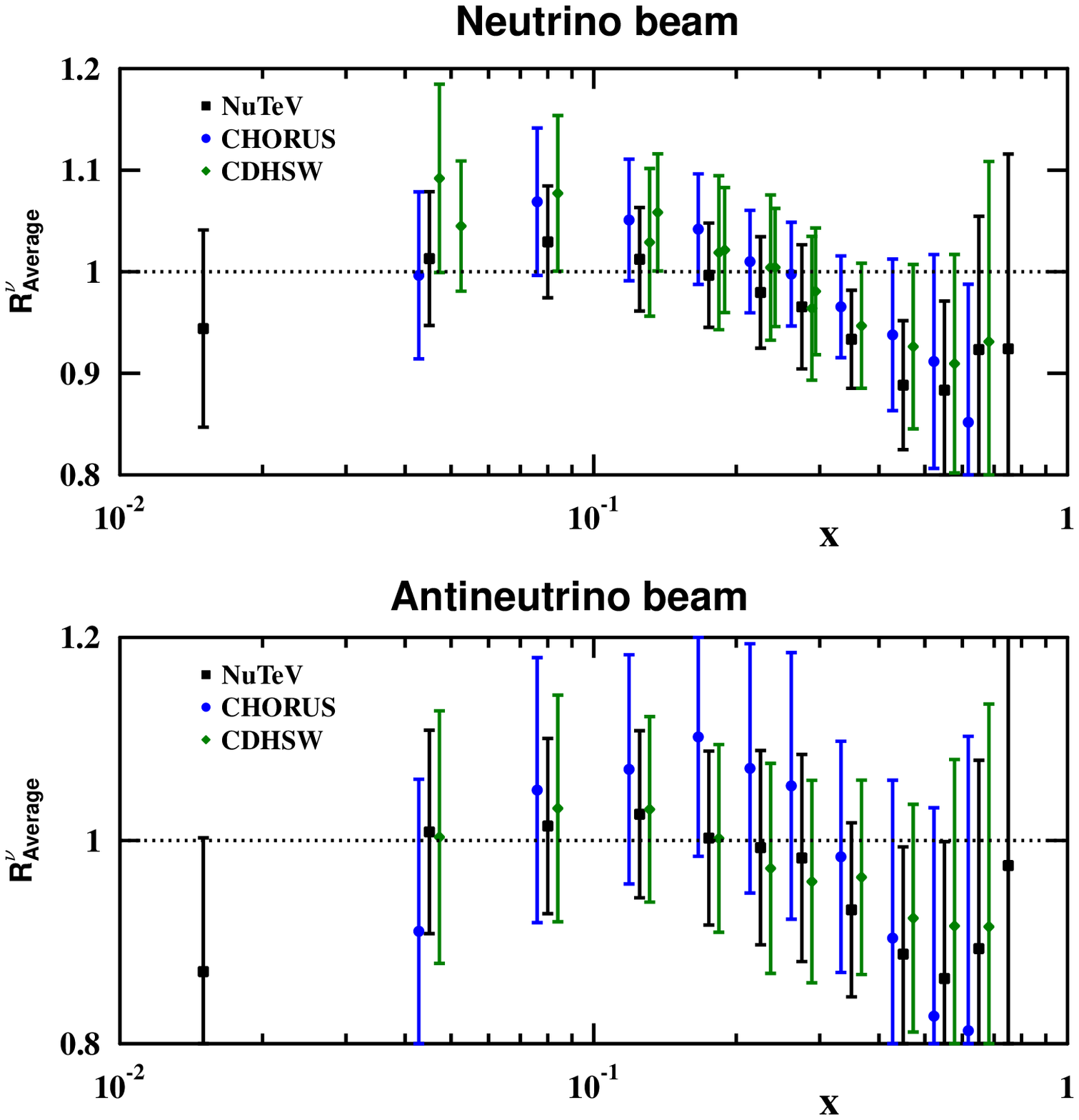}
\end{minipage}
\hspace{0.5cm}
\begin{minipage}[b]{0.45\linewidth}
\centering
\includegraphics[width=\textwidth]{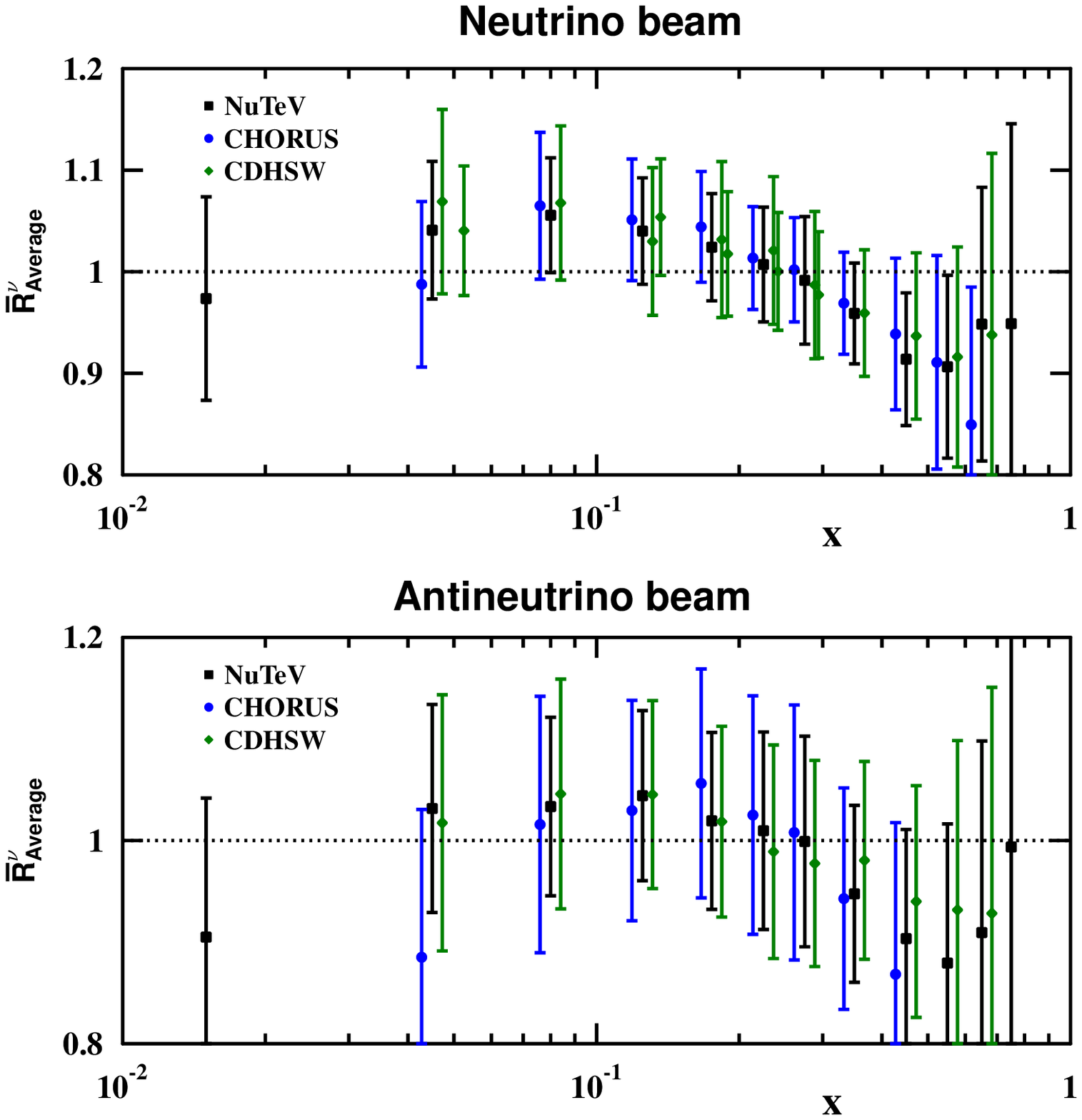}
\end{minipage}
\caption{The neutrino and antineutrino data presented as $R^{\nu}_{\rm Average}$ (left-hand panels),
and as $\overline R^{\nu}_{\rm Average}$ (right-hand panels). The CHORUS (blue circles) and CDHSW (green diamonds)
data have been horizontally shifted from the NuTeV (black squares) data points for clarity.}
\label{Fig:Neutrino1}
\end{figure*}

We utilize the neutrino-nucleus DIS data from the NuTeV \cite{Tzanov:2005kr}, CHORUS \cite{Onengut:2005kv}
and CDHSW \cite{Berge:1989hr} experiments. The difficulty in dealing with the neutrino data is that no
reference data from hydrogen or deuterium target are available and we are forced to use the absolute
experimental cross sections $\sigma^{\nu}_{\rm exp}$ instead of cross section ratios. However, in order
to better see the nuclear effects we still prefer to present the data as ratios
\begin{equation}
 R^{\rm \nu}(x,y,E) \equiv \frac{\sigma^{\nu}_{\rm exp}(x,y,E) }{\sigma^{\nu}_{\rm CTEQ6.6}(x,y,E)}, \label{eq:StandardR}
\end{equation}
where the theoretical cross sections ${\sigma^{\nu}_{\rm CTEQ6.6}}$ are calculated with the CTEQ6.6M central set.
As in \cite{Paukkunen:2010hb}, the theoretical calculations include corrections for the target mass and
electroweak radiation, and are carried out in the SACOT-prescription \cite{Kramer:2000hn} of the variable
flavor number scheme. In order to avoid higher-twist effects we restrict the virtuality $Q^2$ and the final state
invariant mass $W$ by conditions
$Q_{\rm cut}^2 > 4 \, {\rm GeV}^2$, and $W_{\rm cut}^2  > 12.25 \, {\rm GeV}^2$.
This leaves us with 2136 NuTeV, 824 CHORUS, and 937 CDHSW data points.
For a concise presentation of this large amount of data, we form an average 
\begin{equation}
 R^{\nu}_{\rm Average}(x) \equiv
 \left( \sum^N_{i\in {\rm fixed} \, x} \frac{R_i^{\nu}}{\delta_i} \right) \left( \sum^N_{i\in {\rm fixed} \, x} \frac{1}{\delta_i} \right)^{-1}
 \pm
 N \times \left( \sum^N_{i\in {\rm fixed} \, x} \frac{1}{\delta_i} \right)^{-1},
\end{equation}
where $\delta_i$ is the experimental error (statistical and systematic added in quadrature),
and the sum runs over all data points in the same $x$ bin.
This procedure neatly summarizes the main features of the neutrino data as a
function of $x$, but we stress that it is used here only for plotting the data,
the numerical results being computed using the absolute cross sections. The ratios constructed
 this way are shown in the left-hand panels of Figure~\ref{Fig:Neutrino1}. Although
the data from different experiments appear to be in rough mutual agreement, the scatter is 
still non-negligible. In particular, the NuTeV neutrino data seem to lie systematically below the rest
and as such are likely to trigger tension in a global fit --- especially so if the NuTeV correlated
systematic errors are taken seriously as in \cite{Kovarik:2010uv} \footnote{The NuTeV data provide the inverse
covariance matrix for calculating the standard $\chi^2$. However,
in order to consistently account for the correlations in what follows, we would need the absolute shift
in every data point due to $1\sigma$ variation of each systematic parameter separately.} . However, as a function of $x$ the \emph{shape}
of the data seems to follow the usual nuclear effect, suggesting that the problem is rather in the
absolute normalization, as already conjectured in \cite{Paukkunen:2010hb}. For this reason, we define
\begin{equation}
 I_{\rm exp}^\nu(E) \equiv \sum_{i\in {\rm fixed} \, E} \sigma_{{\rm exp},i}(x,y,E) \times B_i(x,y),
\end{equation}
and similarly for the theoretical calculation. The factor $B_i(x,y)$ represents the size of the experimental
$(x,y)$ bin making $I_{\rm exp}^\nu(E)$ thereby an estimate for the integrated cross section.
Now, instead of Eq.~(\ref{eq:StandardR}) we consider the ratio of the {\it normalized} cross sections
\begin{equation}
 \overline R^{\rm \nu}(x,y,E) \equiv \frac{\sigma^{\nu}_{\rm exp}(x,y,E)/I^\nu_{\rm exp}(E)}{\sigma^{\nu}_{\rm CTEQ6.6}(x,y,E)/I^\nu_{\rm CTEQ6.6}(E)}.
  \label{eq:SNewR}
\end{equation}
The averaged neutrino and antineutrino data normalized in this way are plotted in the right-hand panels
of Figure~\ref{Fig:Neutrino1}, demonstrating how all the considered data seem to fall in agreement.
In particular, the NuTeV neutrino data have moved upwards while the CHORUS and CDHSW
neutrino data have remained essentially unchanged.
This observation suggests that the origin of the difficulties in accommodating
the neutrino data in a global fit \cite{Kovarik:2010uv} is due to an unnoticed
problem in the experimental normalization of the NuTeV data --- that the 
uncertainties have probably been underestimated by the experiment. 

\begin{figure*}[ht]
\begin{minipage}[b]{0.45\linewidth}
\centering
\includegraphics[width=\textwidth]{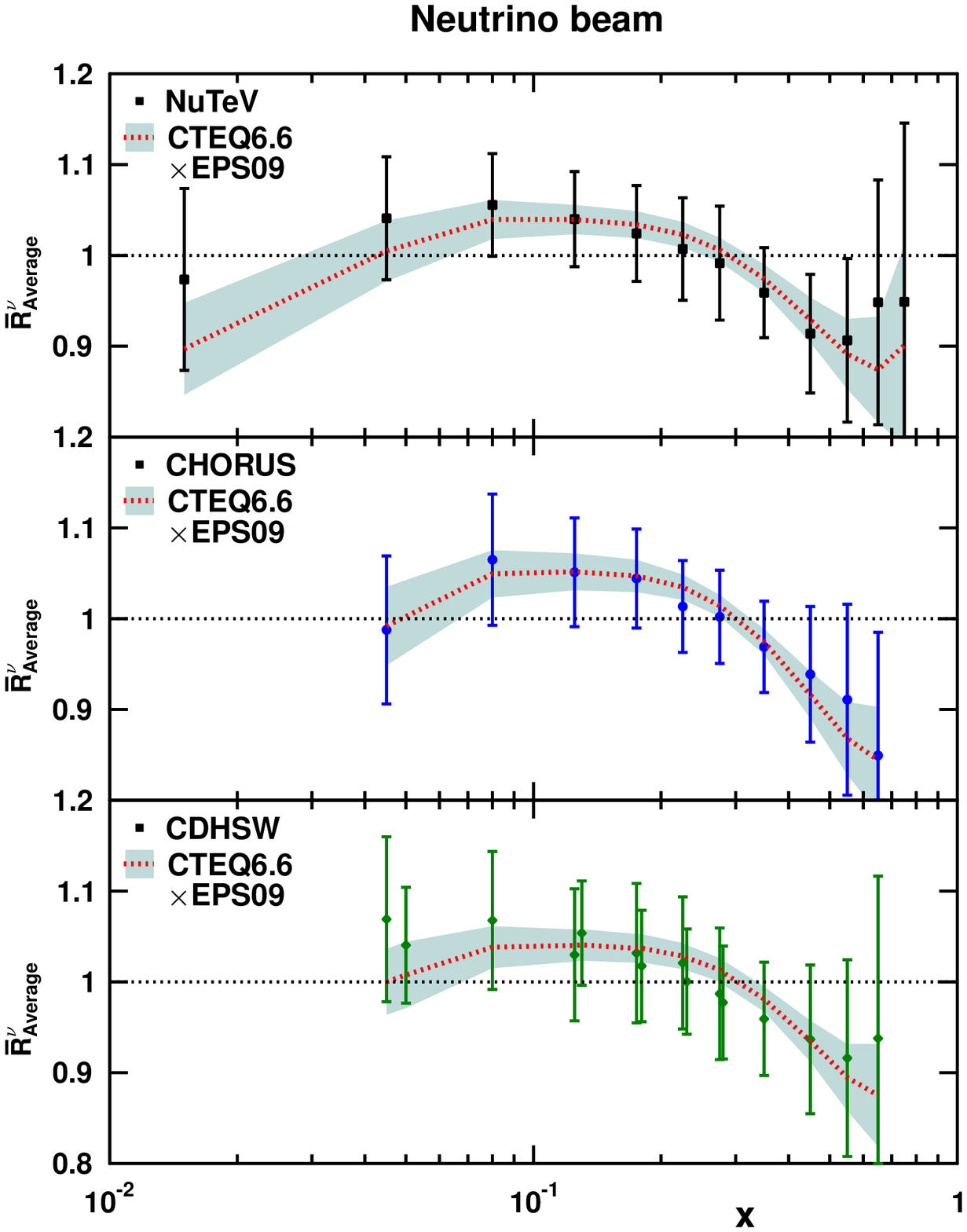}
\end{minipage}
\hspace{0.5cm}
\begin{minipage}[b]{0.45\linewidth}
\centering
\includegraphics[width=\textwidth]{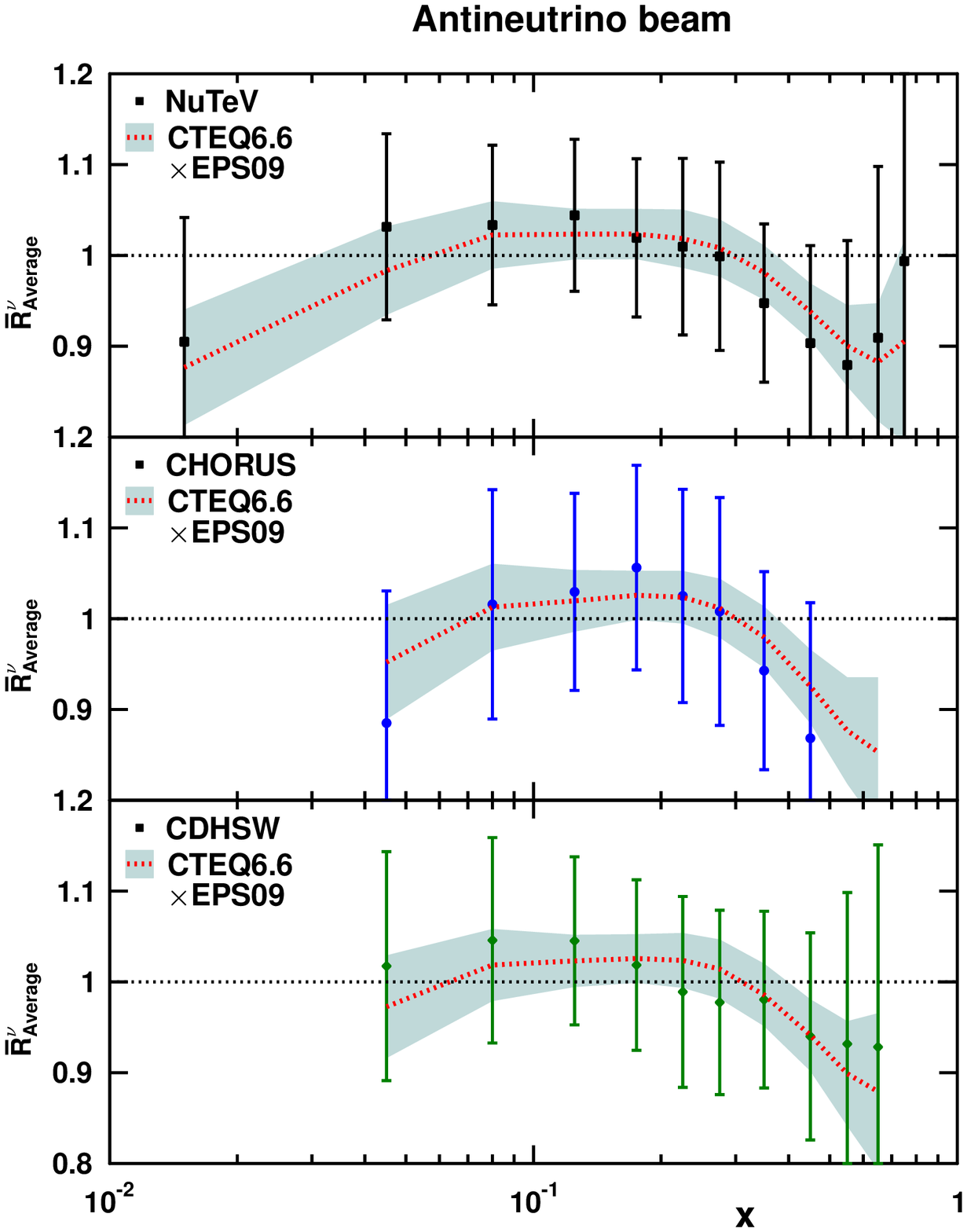}
\end{minipage}
\caption{The experimental $\overline R^{\nu}_{\rm Average}$ compared to the predictions from CTEQ6.6 and EPS09.
}
\label{Fig:Neutrino3}
\end{figure*}

In order to see how the normalized data compare with the predictions from the 
present nuclear PDFs, we replace in Eq. (\ref{eq:SNewR}) the experimental cross sections by the theoretical
ones computed with the bound proton PDFs $f_{i}^A(x,Q^2)$ obtained standardly by
\begin{equation}
f_{i}^A(x,Q^2) \equiv R_{i}^{A, {\rm EPS09}}(x,Q^2) f_{i}^{\rm CTEQ6.6M}(x,Q^2), \label{eq:nPDF}
\end{equation}  
where the factor $R_{i}^{A, {\rm EPS09}}$ represents the EPS09 \cite{Eskola:2009uj} 
nuclear modification in free proton PDF $f_{i}^{\rm CTEQ6.6M}(x,Q^2)$.
The results are shown in Figure~\ref{Fig:Neutrino3}, where
the data points are the same as in the right-hand panels of Figure~\ref{Fig:Neutrino1}, and
the band indicates the theoretical calculations with all PDF uncertainties added
in quadrature \cite{Eskola:2009uj}.
The good agreement indicates that it should be possible to include these data in
global fits without significant mutual disagreement or tension with the other data sets. We note that in the normalization
procedure described here, also part of the PDF uncertainties cancel thereby making the theoretical predictions more solid.

We turn now to a more quantitative description of the data sets accounting for the normalization.
The technique described here is based on the Hessian uncertainty analysis \cite{Pumplin:2001ct}
performed e.g. in the EPS09 and CTEQ6.6 global fits \footnote{The method described here has 
its counterpart in the NNPDF reweighing technique \cite{Ball:2010gb}.}. The neighborhood of the minimum
$\chi^2$ is approximated by an expansion
\begin{equation}
 \chi^2 \approx \chi^2_0 + \sum_{ij} \delta a_i H_{ij} \delta a_j = \chi^2_0 + \sum_i  z_i^2, \label{eq:origchi2}
\end{equation}
where $\delta a_j$ is the deviation of the fit parameter $a_j$ from its best-fit value.
By diagonalizing the Hessian matrix $H_{ij}$ one finds the uncorrelated parameter directions $z_j$
in terms of which the central set $S_0$, and the error sets $S_k^\pm$ are defined:
\begin{eqnarray}
 z(S_0) & = & \left(0,0,...,0 \right) \nonumber \\
z(S^\pm_1) & = & \pm \sqrt{\Delta \chi^2} \left(1,0,...,0 \right) \\
z(S^\pm_2) & = &  \pm \sqrt{\Delta \chi^2} \left(0,1,...,0 \right) \nonumber \\
         & \vdots & \nonumber
\end{eqnarray}
where $\Delta \chi^2$ is the maximum permitted deviation from the minimum $\chi^2$.
These sets enable the calculation of any PDF-dependent quantity $X$ at
the origin and at the corners of the $z$ space, but in order to obtain an
estimate in an arbitrary point $z(S)=(z_1,z_2,\ldots)$ close to the origin, we need to use a linear approximation
\begin{equation}
 X \left[S \right] \approx X \left[S_0 \right] + \sum_k \frac{\partial X \left[S \right]}{\partial z_k}{\Big|_{S=S_0}} z_k
                   \approx X_0 + {\bf D} \cdot {\bf w}, \label{eq:XS}
\end{equation}
where
\begin{eqnarray}
D_k & \equiv & \frac{X\left[S_k^+ \right] - X\left[S_k^- \right]}{2} \\
w_k   & \equiv & \frac{z_k}{\sqrt{\Delta \chi^2}}.
\end{eqnarray}
Let us now consider a larger data set $\{X^{\rm data}\}$. The agreement 
with the PDF set $S$ can be quantified by formally adding its $\chi^2$ contribution $\chi^2\{{X^{\rm data}}\}$ to Eq.~(\ref{eq:origchi2})
\begin{equation}
 \chi^2 = \chi^2_0 + \sum_{\{X^{\rm data}\}} \left[ \frac{X_k\left[S\right] - X_k^{\rm data}}{\delta_k^{\rm data}} \right]^2 + \Delta \chi^2 \sum_k w_k^2, \label{eq:newchi2}
 \end{equation}
where $\delta_k^{\rm data}$ is again the experimental uncertainty and each $X_k\left[S\right]$ is given by Eq.~(\ref{eq:XS}).
The weight vector ${\bf w}$ that minimizes the above $\chi^2$ is given by $ {\bf w}_{\rm min} = -{\bf B}^{-1}  {\bf a}$, where
\begin{eqnarray}
 B_{ij} & = & \sum_k \frac{D^k_i D^k_j}{\left(\delta_k^{\rm data} \right)^2} + \Delta \chi^2 \delta_{ij} \\
 a_i    & = & \sum_k \frac{D_i^k \left( X_k\left[S_0\right] - X_k^{\rm data} \right)}{\left(\delta_k^{\rm data} \right)^2} \\
 D_l^k  & = & \frac{X_k\left[S_l^+ \right] - X_k\left[S_l^- \right]}{2}.
\end{eqnarray}
The level of agreement between the data set $\{X^{\rm data}\}$ and the given set of PDFs is 
now quantified --- not only by $\chi^2\{{X^{\rm data}}\}$  --- but also by the length
of the weight vector ${\bf w}_{\rm min}$. If $|{\bf w_{\rm min}}| < 1$ the new data set could be included
to the original fit within the confidence criterion determined in the analysis.
That is, the ``penalty term'' $\Delta \chi^2 \sum_k w_k^2$ in Eq.~(\ref{eq:newchi2})
remains below the acceptable value $\Delta \chi^2$. On the other hand, if $|{\bf w}_{\rm min}| > 1$,
notable tension between the new and the old data is bound to exist.

\begin{table*}
\begin{center}
{\footnotesize
\begin{tabular}{rcccc||cc}
& \multicolumn{4}{c||}{All CTEQ6.6 and EPS09 error sets} & \multicolumn{2}{c}{Only EPS09 error sets} \\
&       &           &           &       &            &     \\  
NuTeV \vline & $\chi^2_{w=0}/N$ & $\chi_{w_{\rm min}}^2/N$ & EPS09-penalty & CTEQ-penalty &  $\chi_{w_{\rm min}}^2/N$ & EPS09-penalty \\
Normalization \vline &  0.84     &    0.77        &     13.9      &    35.4      &    0.81        &     33.8 \\
No normalization \vline &  1.04     &    0.90        &     40.3      &    42.5   &    0.94        &     77.4 \\
 &       &           &           &       &            &     \\  
CHORUS \vline & $\chi^2_{w=0}/N$ & $\chi_{w_{\rm min}}^2/N$ & EPS09-penalty & CTEQ-penalty  & $\chi_{w_{\rm min}}^2/N$ & EPS09-penalty \\
Normalization \vline &  0.70     &    0.69        &     2.13      &    2.63     &    0.70        &     2.48  \\
No normalization \vline &  0.86     &    0.81        &     3.35      &    14.4    &    0.84        &     5.13  \\
&       &           &           &       &            &     \\  
CDHSW \vline & $\chi^2_{w=0}/N$ & $\chi_{w_{\rm min}}^2/N$ & EPS09-penalty & CTEQ-penalty & $\chi_{w_{\rm min}}^2/N$ & EPS09-penalty \\
Normalization \vline &  0.70     &    0.64        &     7.20      &    17.3      &    0.68        &     9.26 \\
No normalization \vline &  0.81     &    0.74        &     10.4      &    17.8     &    0.78        &     14.1 \\ \\
\end{tabular}
}
\caption[]{\small The $\chi^2/N$ values for the neutrino data and the penalties induced in EPS09 and CTEQ6.6.
The left-hand block of the table corresponds to the analysis including both EPS09 and CTEQ6.6 error sets, while
the right-hand block corresponds to keeping the free proton PDFs fixed at their central value.
}
\label{Table:Data}
\end{center}
\end{table*}

Applying this method to the neutrino data we use the nuclear PDFs defined in Eq.~(\ref{eq:nPDF})
to calculate the cross sections. Therefore, the penalty term splits in two pieces
\begin{equation}
\Delta \chi^2 \sum_k w_k^2 \rightarrow \Delta \chi^2_{\rm EPS09} \sum_{k=1}^{15} w_k^2 + \Delta \chi^2_{\rm CTEQ6.6} \sum_{k=16}^{37} w_k^2
\end{equation}
where  $\Delta \chi^2_{\rm CTEQ} = 100$, and EPS09 $\Delta \chi^2_{\rm EPS09} = 50$.
The key results are given in Table~\ref{Table:Data}
for each data set separately. The $\chi^2_{w=0}$ is the value calculated with the central sets,
whereas $\chi^2_{w_{\rm min}}$ is the corresponding value at the minimum
of Eq.~(\ref{eq:newchi2}). The penalty columns indicate the growths induced in EPS09 and CTEQ6.6 and
the results are given with and without the normalization procedure of Eq.~(\ref{eq:SNewR}).	

The left-hand block of Table~\ref{Table:Data} corresponds to the full analysis
with all EPS09 and CTEQ6.6 error sets. As expected, the
normalization improves the $\chi^2$ values and diminishes the induced
penalties which clearly stay within the allowed range. That is, the
normalized neutrino data could be included
in these global fits without an obvious disagreement with the other data. However, without
the normalization the NuTeV data induce a penalty in EPS09 which starts to get
close to the upper limit $\Delta \chi^2_{\rm EPS09}=50$. Indeed, had we 
taken the free proton PDFs as fixed ($w_k=0$ for $k=16\ldots37$) as in \cite{Kovarik:2010uv}, the EPS09 penalty would
have been much larger. This is demonstrated in the right-hand part of Table~\ref{Table:Data}: Whereas the CHORUS
and CDHSW data stay well inside the permitted region, the the NuTeV data now
cause excess penalty in EPS09. That is, there would be a possible contradiction.

In conclusion, we have demonstrated that disposing the overall normalization
by dividing the data by the integrated cross section in each
neutrino energy bin separately, all large-$Q^2$ neutrino data show practically
identical nuclear effects, consistent with the present nuclear
PDFs. Our numerical consistency test based on the Hessian method of propagating 
uncertainties confirms that these data could be included in a global fit
without causing disagreement with the other data.

In contrast, without the normalization procedure the nuclear effects
preferred by different data sets become much more scattered. In particular,
the NuTeV data seem to
display tension with the other data. Such is not completely unexpected
as in Ref. \cite{Paukkunen:2010hb} sizable differences in the normalization
of the NuTeV data among different neutrino energy bins were found.
This likely explains the findings of \cite{Kovarik:2010uv} where,
however, all neutrino data were rejected as incompatible. The 
analysis reported here suggests that such a strong conclusion is not
justified, and we propose a method to deal with the apparent tension in 
different data sets so that the neutrino data can safely be used in global fits.

\section*{Acknowledgments}
H.P. is supported by the Academy of Finland, Project No. 133005. C.A.S.  is supported by European
Research Council grant HotLHC ERC-2011-StG-279579, by Ministerio de Ciencia e Innovaci\'on of Spain under
grant No. FPA2009-06867-E, and by Xunta de Galicia.

\end{document}